\RequirePackage{fix-cm}
\documentclass[twocolumn,epjc3]{svjour3}
\smartqed  
\RequirePackage{graphicx}
\journalname{}
\begin{document}

\title{Does Dark Matter admixed pulsar exist ?
}


\author{Sajahan Molla $^{1,a}$
        \and
        Bidisha Ghosh$^{2,b}$
        \and
        Mehedi Kalam$^{2,c}$
}

\thankstext{e1}{e-mail: sajahan.phy@gmail.com}
\thankstext{e2}{e-mail: bidishaghosh.physics@gmail.com}
\thankstext{e3}{e-mail: kalam@associates.iucaa.in}


\institute{Department of Physics, New Alipore College, L Block,
New Alipore, Kolkata - 700 053, India \and Department of Physics,
Aliah University, IIA/27, Action Area II, Newtown, Kolkata
-700156, India}

\date{Received: date / Accepted: date}

\maketitle

\begin{abstract}
In this paper, we have considered a two-fluid model assuming that
the pulsars are made of ordinary matter admixed with dark
matter.Contribution of dark matter comes from the fitting of the
rotation curves of the SPARC sample of galaxies\cite{Lelli2016}.
For this we have investigated the dark matter based on the
Singular Isothermal Sphere (SIS) dark matter density profile in
the galactic halo region. Considering this two-fluid model, we
have studied the physical features of the pulsars present in
different galaxy in details. Here, we compute the probable radii,
compactness (u) and surface red-shift ($Z_{s}$) of the four
pulsars namely : PSR J1748-2021B in NGC 6440B galaxy, PSR
J1911-5958A in NGC 6752 galaxy, PSR B1802-07 in NGC 6539 galaxy
and PSR J1750-37A in NGC 6441 galaxy. 
\end{abstract}

\section{Introduction}
\label{intro} It was noticed that the study of compact objects
take much attention to the astro physicist during the last few
decade due to their unique properties compare to an Earth-based
experiments. Compact objects comprise excellent natural
laboratories to study, test and constrain new physics and/or
alternative theories of gravity under extreme conditions. Compact
relativistic objects such as white dwarfs, neutron stars and
black hole are the last destiny of the evolved stars
\cite{Shapiro1983}. The stars become stable when the outward
degeneracy pressure provided by the Fermi gas balances the inward
gravitational force. In case of white dwarfs the Fermi gas
consists of electrons while in neutron stars Fermi gas consists
of neutrons. Normally, neutron stars are composed mostly by
neutrons while a new object, called strange stars are made of
strange quark matter (SQM) or its conversion (u,d,s quarks) and
they may be enclosed to the core of the neutron star
\cite{Drago2014,Haensel1986}. It is familiar that neutron stars
are bounded by gravitational attraction where as strange stars
are bounded by strong interactions as well as gravitational
attractions. Therefore, strange stars become more gravitationally
bound than neutron stars. Since a strange star is more stable
compared to a regular neutron star, its formation could interpret
the origin of the huge amount of energy released in superluminous
supernovae \cite{Leahy2008}. This type of supernovae event
happens about one out of every 1000 supernovae explosions and it
is more than 100 times brighter than common supernovae. In most
cases, a strange star and a neutron star can be separated on the
basis of their vanishing surface energy density
\cite{Haensel1986,Alcock1986,Farhi1984,Postnikov2010,Dey1998}.
Since after the birth of a neutron star, within a few second, its
temperature becomes less than the Fermi energy, hence, for a given
equation of state the mass and radius of the neutron star depend
only on central density. Although, it is very difficult to find
it's mass and radius simultaneously. We suggest to see a review
work of Lattimer $\&$ Prakash 2007 \cite{lattimer2007} for a
detail study. From the solutions of Tolman-Oppenheimer-Volkoff
equations, we can theoretically enumerate the mass and radii of
the spherically-symmetric compact stars.The mass and radius of a
compact star can be measured by pulsar timing, thermal emission
from cooling stars, surface explosions and gravity wave emissions
through observations.It is well known that the properties of the
compact objects like mass and radius, crucially depend on the
equation of state, unfortunately, which is poorly known to us.
Truly, the most challenging task is to fix the exact Equation of
State (EoS) to describe the structure of a compact star
\cite{Ozel2006,Ozel2009,Ozel2009a,Ozel2010,Guver2010,Guver2010a}.
Though few compact star's masses have been decided (to some
extend), which are in binaries
\cite{Heap1992,lattimer2005,Stickland1997,Orosz1999,Van1995} but
there is no information about the radius. Therefore,the
theoretical study of the stellar structure is required to support
the correct direction for the newly observed stelar masses. In
these ground some of the researcher's work on compact stars has
been mentioned here \cite{Rahaman2012a,Rahaman2012b,Kalam2012a,Kalam2012b,Kalam2013,Kalam2014a,Kalam2014b,Kalam2016,Kalam2017,Jafry2017,Hossein2012,Lobo2006,Bronnikov and Fabris2006,Maurya2016,Dayanandan2016,Maharaj2014,Ngubelanga2015,Paul2015,Pant2014,Piyali2017}.

In 1933 Zwicky discovered the dark matter while studying the
dynamic properties of the Coma galaxy cluster \cite{Zwicky1933}.
Whereas Rubin and Ford arrived at the similar conclusions about
the existence of the dark matter with optical studies of galaxies
like M31 \cite{Rubin1970} after few decades. We suggest to see
the Ref. \cite{Olive2004,Munoz2004}, a review work on dark
matter. In fact, we are still unaware of the origin and nature of
the dark matter. In reality, the kind of elementary particles
playing the role of dark matter in the universe is one of the
recent challenges of particle physics and modern cosmology. After
an intensive study, cosmologists and particle physicists recently
proposed many dark matter candidates to explain or constrain the
properties
\cite{Taoso2008,Lopes2010,Kouvaris2010,Turck2012,Lopes2014,Lopes2014a,Brito2015,Brito2016,Martins2017}.

Though dark matter does't interact directly with normal matter, it
can have significant gravitational effects on stellar objects
\cite{Li2012a,Li2012b,Panotopoulos2017a}. It was reported that
\cite{Narain2006,Leung2011,Leung2012,Mukhopadhyay2016,Panotopoulos2017b}
the fermionic dark matter can have more gravitational effects on
strange star's physical properties. Till today the spin of the
dark matter particles remain unknown to us, one can think of the
bosonic dark matter scenario. The authors Panotopoulos and Lopes
\cite{Panotopoulos2017} studied the effects of bosonic condensed
dark matter on strange stars as far as the radial oscillations
are concerned. It is generally accepted that dark matter
particles are collisionless. Spergel and Steinhardt
\cite{Spergel2000} inaugurated the idea that dark matter may have
self-interaction in order to wipe out some apparent conflicts
among the collisionless cold dark matter example and
astrophysical observations. The dark matter core inside the
neutron star considered as continuity in Ref. \cite{Ellis2018},
whereas in Ref. \cite{Nelson2018}, it was considered that dark
matter halo enveloping the star. It was reported that in the
presence of dark matter core scenario, possible effects on the
maximum mass of a neutron star, its radius for any fixed mass,
and its tidal deformability $\Lambda$ are in general reduced
\cite{Panotopoulos2017b,Ellis2018a}. Whereas an increase in tidal
deformability $\Lambda$ was found in the dark matter halo model
scenario \cite{Nelson2018}.\\


Generally Pulsar is a subclass of Neutron stars and it became an
interesting object to the astrophysicists for the last few years.
Freire et al. \cite{Freire2008} measured the mass of the pulsar
"PSR J1748-2021B" in NGC 6440B by using the Green Bank
Telescope's S-band receiver and the Pulsar Spigot spectrometer
and it comes out as $2.74 \pm 0.21 M_{\odot}$. In another work,
Bassa et al. \cite{Bassa2006} present spectroscopic and
photometric observations of the optical counterpart to PSR
J1911-5958A, a millisecond pulsar located in the globular cluster
NGC 6752 and it's mass comes out as $1.40^{+0.16}_{-0.10}
M_{\odot}$. On the other hand, S.E. Thorsett and D. Chakrabarty
\cite{Thorsett1999} reported the measurement of the mass of the
PSR B1802-07 which is in the globular cluster NGC 6539, as
$1.26^{+0.08}_{-0.17} M_{\odot}$. Freire et al.
\cite{Lattimer2012} have also determined the mass of the pulsar
"PSR J1750-37A" in NGC 6441 as $1.26^{+0.39}_{-0.36} M_{\odot}$.


The Singular Isothermal Sphere (SIS) model is very useful for
gravitational lensing\cite{Keeton2002}.Also, so many
observational evidences are there where they are consistent with
SIS profiles.This motivates us to study the Singular Isothermal
Sphere(SIS) model\cite{Keeton2002} to study the dark matter.

Also, we have been motivated by the previous articles on dark
matter neutron star reported in well known journals
\cite{Li2012b,Leung2011,Panotopoulos2017b,Sandin2009,Goldman2013,Leung2013,Mukhopadhyay2017,Rezaei2017,Rezaei2018}.
The motto of the present article is to study the existence of dark
matter with ordinary matter in pulsars located in various
galaxies, namely, PSR J1748-2021B in NGC 6440B, PSR J1911-5958A
in NGC 6752, PSR B1802-07 in NGC 6539, PSR J1750-37A in NGC 6441.
We have come to a conclusion that there is every possibility of
existence of dark matter admixed with ordinary matter in the
above mentioned pulsars.

\section{Interior Spacetime}
\label{sec:1}
Consider a static and spherically symmetric star with interior spacetime as
\begin{equation}
ds^2=-e^{\nu(r)}dt^2 + e^{\lambda(r)}dr^2 +r^2(d\theta^2 +sin^2\theta d\phi^2) \label{eq1}
\end{equation}
Inspired from many previous revealed famous articles \cite{Takisa2019,Matondo2018,Maurya2015,Maurya2019,Dayanandana2017,Maurya2017,Gedela2018,Singh2017,Rahaman2017,Jasim2016,Takisa2016,Maurya2016,Singh2017a}, we have taken static, spherical symmetric metric for this pulsar model.
The Einstein field equations for the metric Eq.~(1) obtained as (taking $G = c = 1$)
\begin{eqnarray}
8\pi \rho &=& e^{-\lambda}\left(\frac{\lambda^\prime}{r}-\frac{1}{r^2}\right) + \frac{1}{r^2} \label{eq2} \\
8\pi p_r &=& e^{-\lambda}\left(\frac{\nu^\prime}{r}+\frac{1}{r^2}\right) - \frac{1}{r^2} \label{eq3} \\
8\pi p_t &=& \frac{e^{-\lambda}}{2}\left(\frac{(\nu^\prime)^2 - \lambda^\prime \nu^\prime}{2} + \frac{\nu^\prime - \lambda^\prime}{r} + \nu^{\prime\prime}\right) \label{eq4}
\end{eqnarray}

According to H. Heintzmann \cite{Heintzmann1969},
\begin{eqnarray}
e^{\nu} &=&
A^2\left(1+ar^2\right)^{3} \nonumber
\end{eqnarray}
and
\begin{eqnarray}
e^{-\lambda} &=&
\left[1-\frac{3ar^2}{2} \left(\frac{1+C\left(1+4ar^2\right)^{-\frac{1}{2}}}{1+ar^2}\right) \right] \nonumber
\end{eqnarray}
Where A (dimensionless), C (dimensionless) and $a$ ( length$^{-2}$ ) are constants (using geometric units G=c=1).\\
Therefore,
\begin{eqnarray}
\nu^\prime &=& \frac{6ar}{1+ar^2} \nonumber
\end{eqnarray}
\begin{eqnarray}
\nu^{\prime\prime} &=& -\frac{6a\left(-1+ar^2\right)}{\left(1+ar^2\right)^2} \nonumber
\end{eqnarray}
\begin{eqnarray}
\lambda^\prime = -\left[{6ar\left(C+2aCr^2-2a^2Cr^4+\left(1+4ar^2\right)^{\frac{3}{2}}\right)}\right] \nonumber \\
\times[\left(1+ar^2\right)\left(1+4ar^2\right) \nonumber \\
\left(-2\sqrt{1+4ar^2}+ar^2\left(3C+\sqrt{1+4ar^2}\right)\right)]^{-\frac{1}{2}} \nonumber
\end{eqnarray}
We also assume that the energy-momentum tensor for the matter distribution of the compact star has the standard form as
\[T_{ij}=diag(-\rho,p_r,p_t,p_t)\]
where $\rho$ is the energy density, $p_r$ and $p_t$ are the radial and transverse pressure respectively. \\

The singular isothermal sphere (SIS) density profile is the
simplest model of the matter distribution in an astrophysical
system. F. Brimioulle et al. \cite{Brimioulle2013} have mentioned
that, dark matter is the dominant part of the galaxy. They have
done  galaxy-galaxy lensing (GGL) study based on imaging data
from the Canada-France-Hawaii Telescope Legacy Survey Wide. They
have fitted three galactic halo profiles to the lensing signal,
among the three, one  is singular isothermal sphere (SIS) dark
matter density profile.

Also, M. Oguri et al. \cite{Oguri2002} have considered dark halo
density profile as the SIS as well as NFW density profile. This
motivates us to take the dark matter density profile as SIS
density profile.

SIS model for dark matter energy density which is applicable to the stellar cores with no nuclear burning i.e. for compact star is as
follows: $\rho_{d} (r)=\frac{K}{2\pi G r^2}$ \& $p_{d} (r)=\frac{mK}{2\pi G r^2}$, where K, m are constant \cite{Barranco2013}.\\

The tangential velocity of the halo region, $v_{halo}^2$ can be
written as \\
\begin{equation}
v_{halo}^2 = \frac{G}{R}\int_0^R 4\pi r^2 \rho_d(r)dr = 2K.
\end{equation}
 \begin{table}
\centering
\begin{center}
\caption{Circular velocity of the flat part of the rotation
curves of galaxies($v_{obs}$), rotational velocity of
halos($v_{halo}$) and the K values of dwarf galaxies which is
calculated from the galactic rotational curve fit of dwarf
galaxies.}
\begin{tabular}{lccc}
\hline \hline
Galaxy &   $v_{obs}^2(\times 10^{-07})$    &  $v_{halo}^2(\times 10^{-08})$  &  $K(\times 10^{-08})$           \\
\hline
NGC 3769           & $ 1.764   $ & $ 9.282   $ & $ 4.641  $   \\
NGC 3877           & $ 3.249   $ & $ 8.853   $ & $ 4.426  $   \\
NGC 3917           & $ 2.116   $ & $ 8.853   $ & $ 4.426  $   \\
NGC 3949           & $ 3.173   $ & $ 4.895   $ & $ 2.447  $   \\
NGC 3972           & $ 1.995   $ & $ 7.567   $ & $ 3.783  $   \\
NGC 4051           & $ 2.880   $ & $ 7.077   $ & $ 3.538  $   \\
NGC 4085           & $ 2.055   $ & $ 2.581   $ & $ 1.290  $   \\
NGC 4088           & $ 3.680   $ & $ 6.032   $ & $ 3.016  $   \\
NGC 4183           & $ 1.469   $ & $ 8.317   $ & $ 4.158  $   \\
NGC 4214           & $ 0.722   $ & $ 4.867   $ & $ 2.433  $   \\
NGC 4217           & $ 4.053   $ & $ 8.650   $ & $ 4.325  $   \\
NGC 7793           & $ 1.495   $ & $ 2.380   $ & $ 1.190  $   \\
NGC 0024           & $ 1.344   $ & $ 7.712   $ & $ 3.856  $   \\
NGC 0300           & $ 1.045   $ & $ 5.471   $ & $ 2.735  $   \\
UGC 00191          & $ 0.781   $ & $ 3.970   $ & $ 1.985  $   \\
UGC 00634          & $ 1.296   $ & $ 9.847   $ & $ 4.923  $   \\
UGC 00731          & $ 0.608   $ & $ 3.282   $ & $ 1.641  $   \\
UGC 00891          & $ 0.452   $ & $ 2.955   $ & $ 1.477  $   \\
UGC 01230          & $ 1.418   $ & $ 6.934   $ & $ 3.467  $   \\
UGC 05716          & $ 0.620   $ & $ 3.756   $ & $ 1.878  $   \\
UGC 05721          & $ 0.758   $ & $ 4.568   $ & $ 2.284  $   \\
UGCA  442          & $ 0.371   $ & $ 2.314   $ & $ 1.157  $   \\
DDO 161            & $ 0.506   $ & $ 2.849   $ & $ 1.424  $   \\
F563-V2            & $ 1.547   $ & $ 3.907   $ & $ 1.953  $   \\
F565-V2            & $ 0.767   $ & $ 4.411   $ & $ 2.205  $   \\
\hline
\end{tabular}
\label{tab:RAR}
\end{center}
\end{table}

\begin{table}
\centering
\begin{center}
\caption{Circular velocity of the flat part of the rotation
curves of galaxies($v_{obs}$), rotational velocity of halos
($v_{halo}$)and the K values of spiral galaxies, which is
calculated from the galactic rotational curve fit of spiral
galaxies.}
\begin{tabular}{lccc}
\hline \hline
Galaxy &   $v_{obs}^2(\times 10^{-07})$    &  $v_{halo}^2(\times 10^{-07})$  &  $K(\times 10^{-07})$           \\
\hline
NGC 3992           & $ 8.220   $ & $ 4.414   $ & $ 2.207  $   \\
NGC 4013           & $ 4.356   $ & $ 2.039   $ & $ 1.019  $   \\
NGC 5033           & $ 5.625   $ & $ 3.427   $ & $ 1.713  $   \\
NGC 2903           & $ 5.184   $ & $ 2.046   $ & $ 1.023  $   \\
NGC 2998           & $ 5.088   $ & $ 3.126   $ & $ 1.563  $   \\
NGC 2841           & $ 11.59   $ & $ 4.443   $ & $ 2.221  $   \\
NGC 2683           & $ 4.993   $ & $ 2.083   $ & $ 1.041  $   \\
NGC 0289           & $ 4.181   $ & $ 2.280   $ & $ 1.140  $   \\
NGC 5371           & $ 6.507   $ & $ 1.801   $ & $ 0.906  $   \\
NGC 5907           & $ 6.136   $ & $ 3.366   $ & $ 1.683  $   \\
NGC 5985           & $ 10.33   $ & $ 4.443   $ & $ 2.221  $   \\
NGC 6195           & $ 7.395   $ & $ 2.146   $ & $ 1.073  $   \\
NGC 6674           & $ 9.409   $ & $ 4.442   $ & $ 2.221  $   \\
NGC 7331           & $ 7.338   $ & $ 2.652   $ & $ 1.326  $   \\
NGC 7814           & $ 7.802   $ & $ 3.562   $ & $ 1.781  $   \\
UGC 02487          & $ 16.29   $ & $ 4.444   $ & $ 2.222  $   \\
UGC 02885          & $ 10.33   $ & $ 4.435   $ & $ 2.217  $   \\
UGC 02953          & $ 11.30   $ & $ 4.444   $ & $ 2.222  $   \\
UGC 03205          & $ 6.240   $ & $ 3.381   $ & $ 1.690  $   \\
UGC 03546          & $ 7.627   $ & $ 2.454   $ & $ 1.227  $   \\
UGC 05253          & $ 6.833   $ & $ 3.076   $ & $ 1.538  $   \\
UGC 06786          & $ 5.826   $ & $ 3.971   $ & $ 1.985  $   \\
UGC 06787          & $ 8.464   $ & $ 4.278   $ & $ 2.139  $   \\
UGC 08699          & $ 4.533   $ & $ 2.093   $ & $ 1.046  $   \\
UGC 09133          & $ 9.280   $ & $ 4.378  $ & $ 2.189    $   \\
\hline
\end{tabular}
\label{tab:RAR}
\end{center}
\end{table}
The tangential velocity at the halo region, $v_{halo}^2$ for dwarf
galaxies are shown in Table 1 and for massive spiral galaxies are
shown in Table 2. It was calculated from the fitting of the
rotation curves of the SPARC sample of galaxies\cite{Lelli2016}.
$v_{halo}^2$ has been calculated from the equation given below :
\begin{equation}
v_{halo}^2 = v_{obs}^2 - \left( v_{star}^2 + v_{gas}^2\right);
\end{equation}

It is to be mentioned here that the K values for both dwarf and
spiral galaxies are of the order of $10^{-7}$ to $10^{-8}$ and
that value of K comes from $v_{halo}^2$,the observational data of
galactic rotational curve. Moreover, this K has an important role
to the density distribution of the dark matter haloes.

Now, we consider the pulsars are made of ordinary matter admixed
with condensed dark matter. Therefore, effective density and
pressure can be written as
\begin{eqnarray}
\rho_{eff} &=& \rho + \rho_{d} \nonumber  \\
p_{eff} &=& p - p_{d} \nonumber
\end{eqnarray}
The presence of dark matter in addition with normal matter the
Einstein field equations for the metric Eq.~(1) can be obtained as
(taking $G = c = 1$)
\begin{eqnarray}
\rho &=& \frac{1}{8\pi}\left[e^{-\lambda}\left(\frac{\lambda^\prime}{r}-\frac{1}{r^2}\right) + \frac{1}{r^2}\right] - \frac{K}{2\pi r^2} \label{eq2} \\
p_{r} &=& \frac{1}{8\pi}\left[e^{-\lambda}\left(\frac{\nu^\prime}{r}+\frac{1}{r^2}\right) - \frac{1}{r^2}\right] + \frac{mK}{2\pi r^2} \label{eq3}
\end{eqnarray}
\begin{equation}
p_{t} = \frac{1}{8\pi}\left[\frac{e^{-\lambda}}{2}\left(\frac{(\nu^\prime)^2 - \lambda^\prime \nu^\prime}{2} + \frac{\nu^\prime - \lambda^\prime}{r} + \nu^{\prime\prime}\right)\right] \nonumber \\
+\frac{mK}{2\pi r^2}
\end{equation}

\begin{figure*}
  \includegraphics[width=0.50\textwidth]{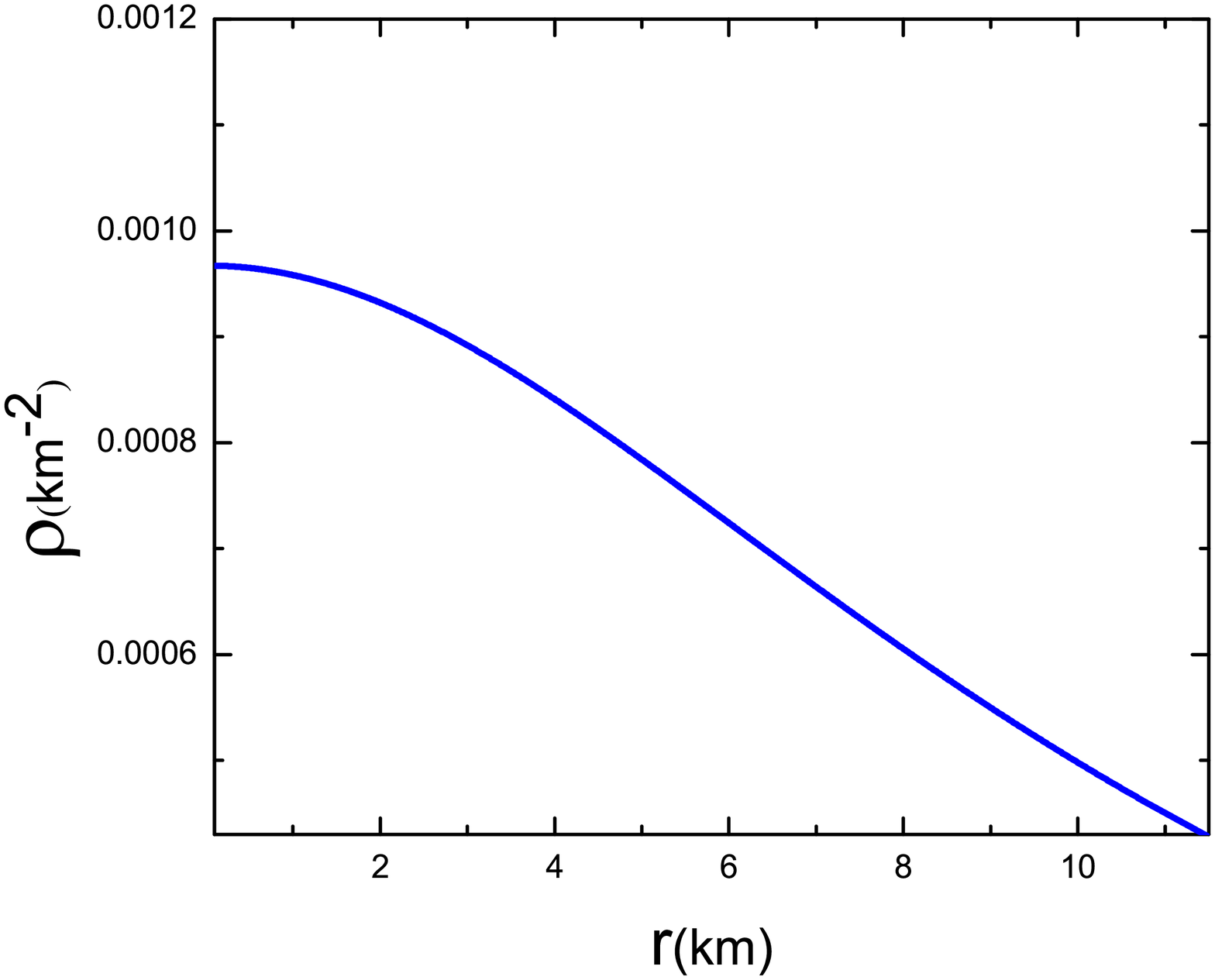}
  \includegraphics[width=0.50\textwidth]{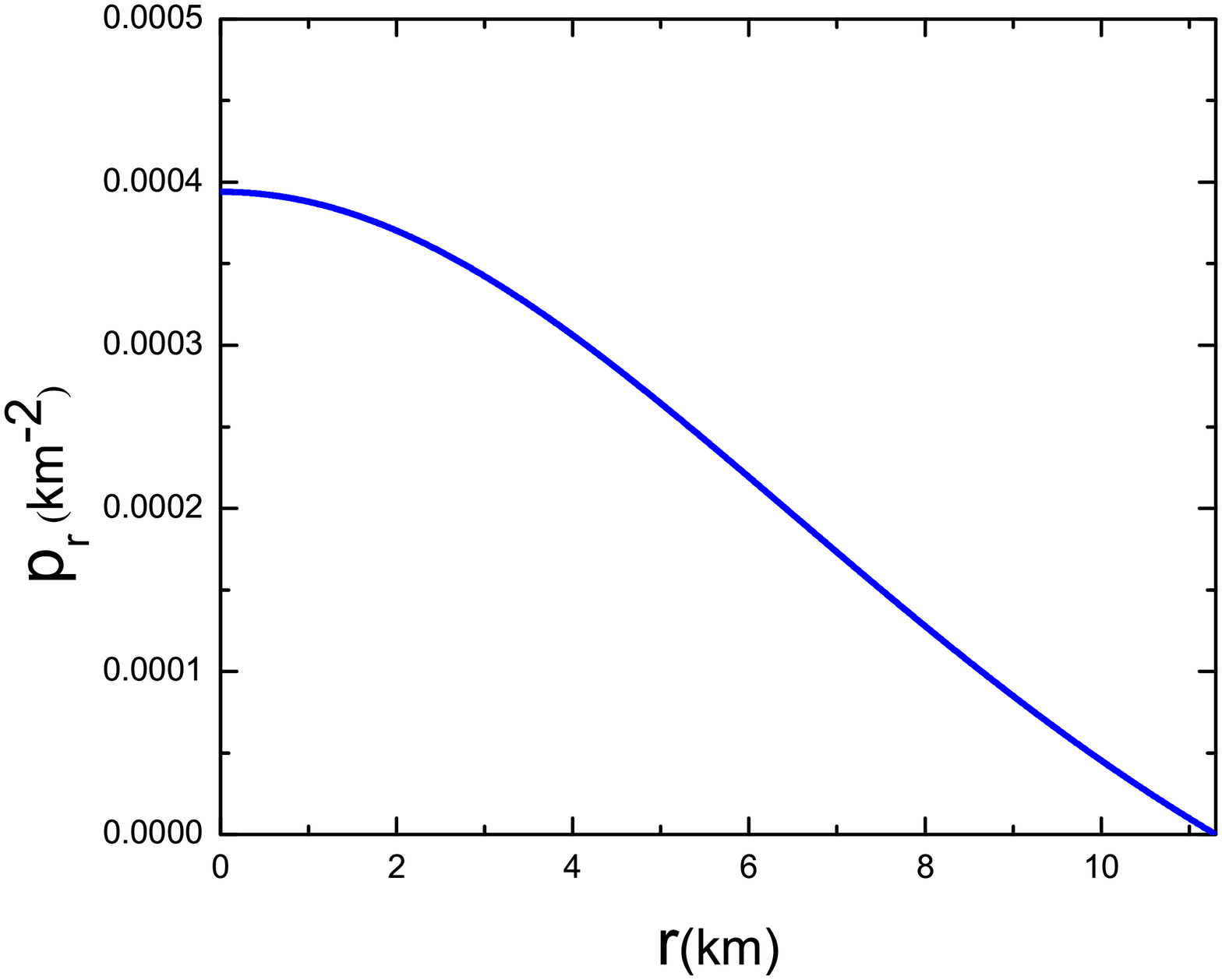}\\
   \includegraphics[width=0.50\textwidth]{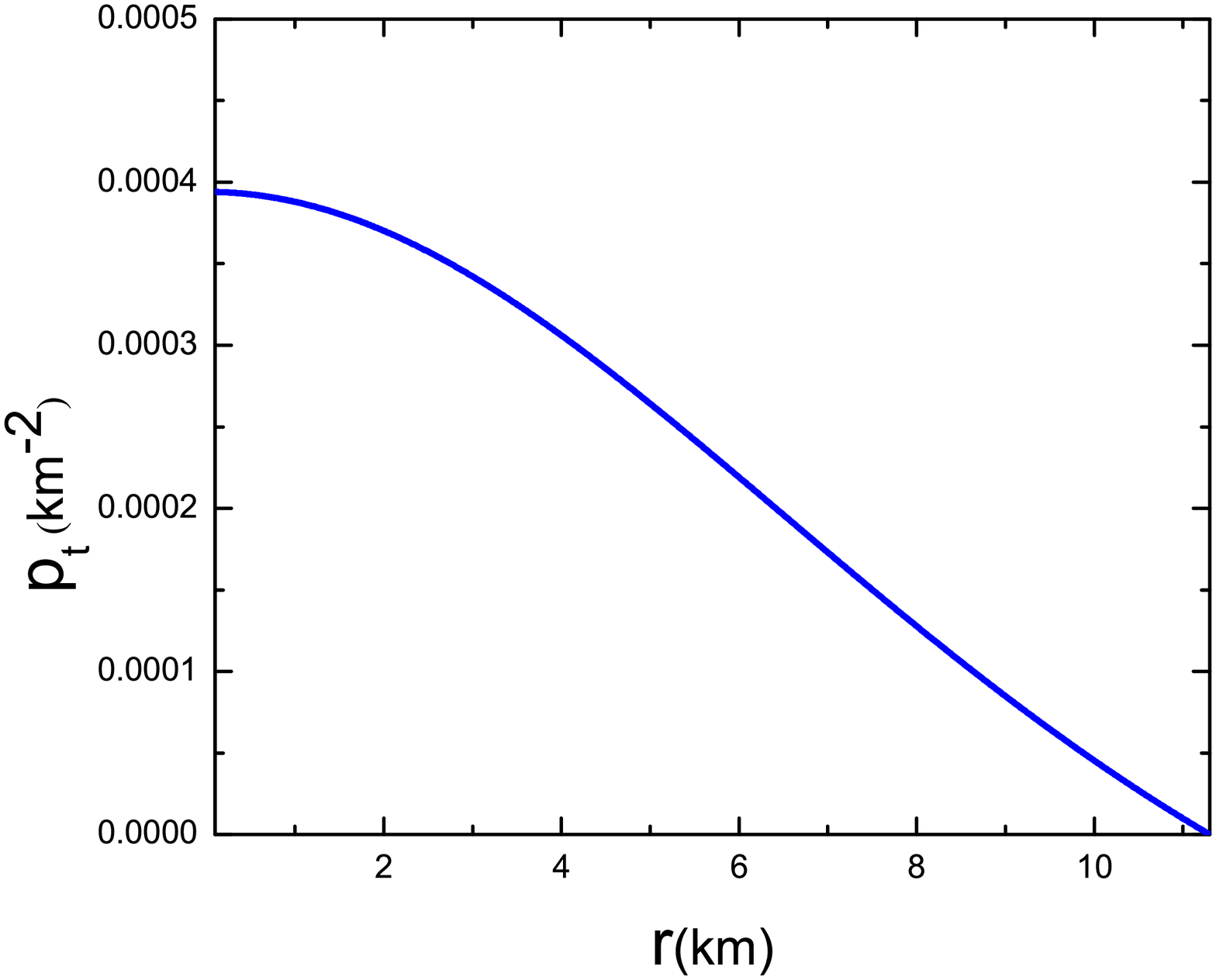}
  \includegraphics[width=0.50\textwidth]{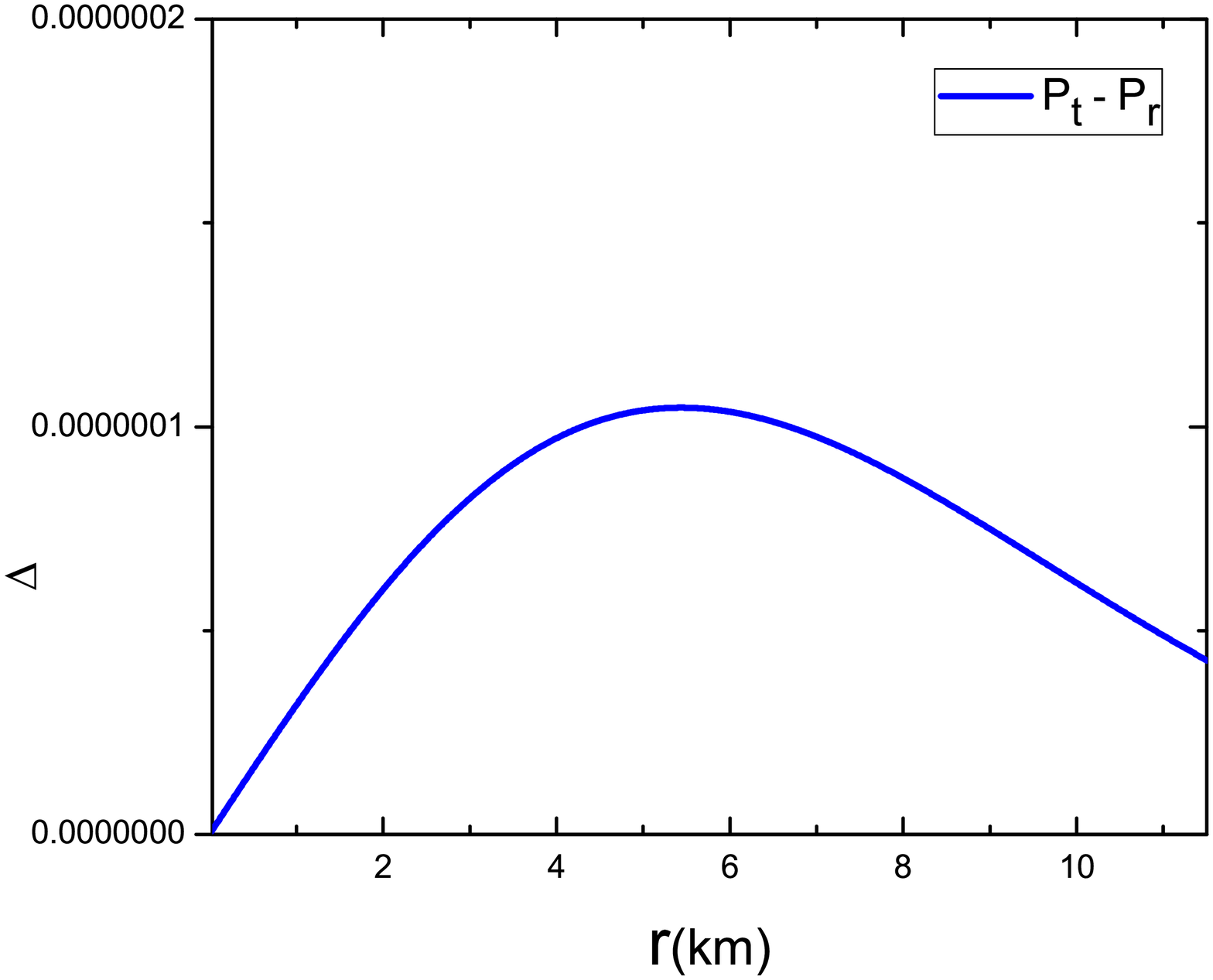}
\caption{Matter density($\rho$) - radius(r) and pressure($p$) -
radius(r) variation at the pulsar interior.} \label{fig:1}
\end{figure*}

Fig.~1 shows that, the matter density and pressure both are
maximum at the centre and decreases monotonically towards the
boundary. Thus, the energy density and the pressure are well
behaved in the interior of the stellar structure. The anisotropic
parameter $\Delta (r) = \left(p_t-p_r\right)$ representing the
anisotropic stress is given by Fig.~1. The `anisotropy' will be
directed outward when $p_{t}>p_{r}$ i.e. $\Delta>0,$ and inward
when $p_{t}<p_{r}$ i.e. $\Delta<0$. It is apparent from the
Fig.~1 of our model that a repulsive `anisotropic' force
($\Delta>0$) allows the construction of more massive
distributions. It is to be mentioned here that, we set the values
of the constants $K=10^{-8}$ ( as observed  $K$ varies from
$10^{-7}$ to  $10^{-8}$), $a=0.003 km^{-2}$, $C=0.8$ and
$m=10^{-6}$, such that the pressure drops to zero at the boundary.

\section{Exploration of Physical properties}
\label{sec:2}
In this section we will investigate the following physical features of the compact object under these model:
\subsection{Energy conditions}
From Fig.~2 we observe that all the energy conditions namely, null energy condition(NEC), weak energy condition(WEC), strong energy condition(SEC) and dominant energy condition(DEC) are satisfied in our stellar model.\\
(i) NEC: $p+\rho\geq0$ ,\\
(ii) WEC: $p+\rho\geq0$  , $~~\rho\geq0$  ,\\
(iii) SEC: $p+\rho\geq0$  ,$~~~~3p+\rho\geq0$ ,\\
(iv) DEC: $\rho > |p| $.

\begin{figure}
\centering
\resizebox{0.50\textwidth}{!}{
  \includegraphics{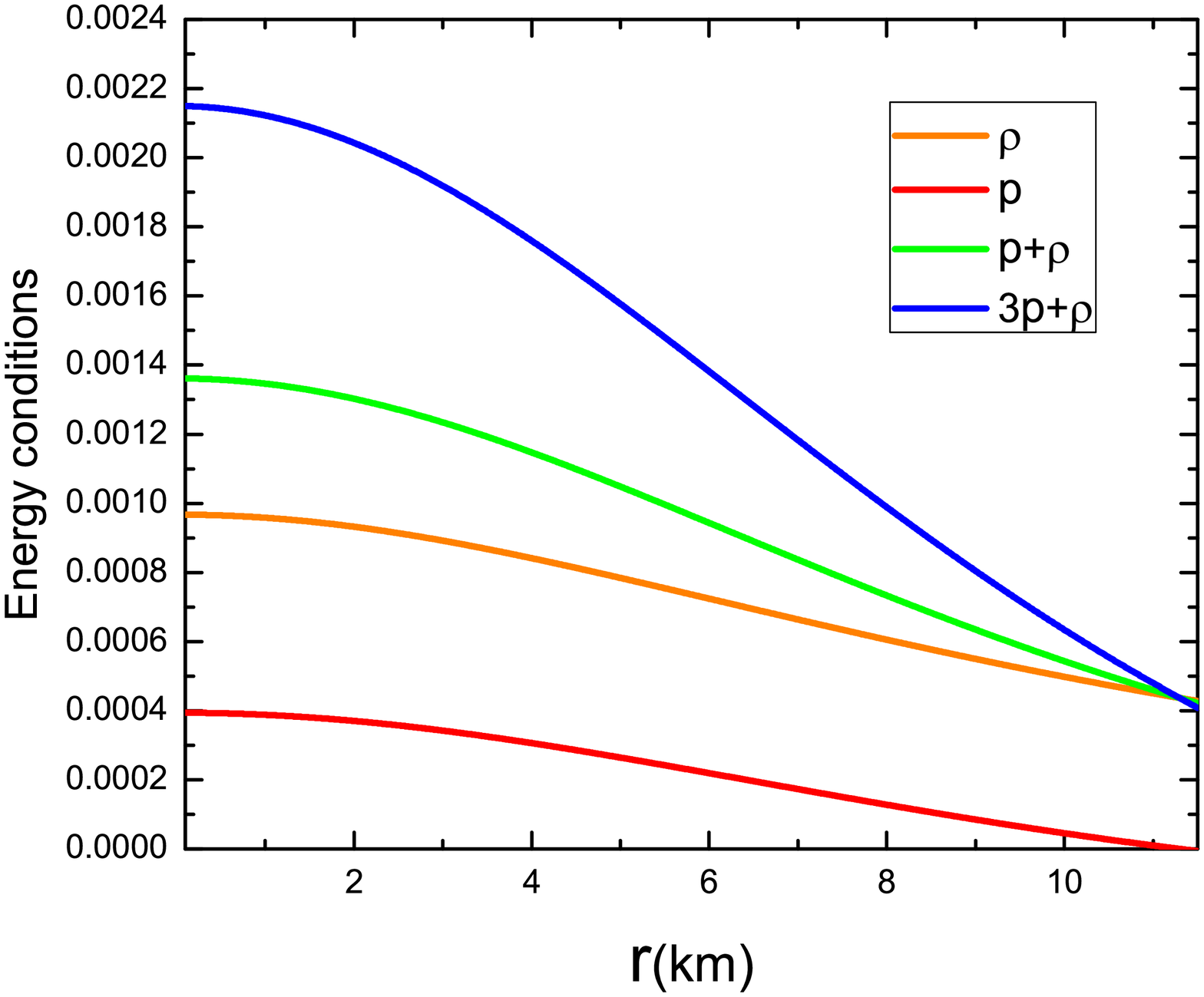}
} \caption{Energy conditions variation at the pulsar interior.}
\label{fig:2}
\end{figure}

\subsection{TOV equation}
For an anisotropic fluid distribution, the generalized TOV equation has the form
\begin{equation}
\frac{dp_{r}}{dr} +\frac{1}{2} \nu^\prime\left(\rho
 + p_{r}\right) + \frac{2}{r}\left(p_{r} - p_{t}\right)
= 0.\label{eq18}
\end{equation}
Following Ref. \cite{Leon1993}, we write the above equation as
\begin{equation}
-\frac{M_G\left(\rho+p_{r}\right)}{r^2}e^{\frac{\lambda-\nu}{2}}-\frac{dp_{r}
}{dr}
 +\frac{2}{r}\left(p_{t}-p_{r}\right) = 0, \label{eq19}
\end{equation}
Where $M_G(r)$ is the gravitational mass inside a sphere of radius $r$ and is given by
\begin{equation}
M_G(r) = \frac{1}{2}r^2e^{\frac{\nu-\lambda}{2}}\nu^{\prime}.\label{eq20}
\end{equation}
which can be derived from the Tolman-Whittaker formula and the Einstein's field equations. The modified TOV equation describes the equilibrium condition for the compact star subject to gravitational ($F_g$) and hydrostatic ($F_h$) plus another force due to the anisotropic ($F_a$) nature of the stellar object as
\begin{equation}
F_g+ F_h + F_a = 0,\label{eq21}
\end{equation}
Where,
\begin{eqnarray}
F_g &=& -\frac{1}{2}\nu^{\prime}\left(\rho+p_{r}\right)\label{eq22}\\
F_h &=& -\frac{dp_{r}}{dr} \label{eq23}\\
F_a &=& \frac{2}{r}\left(p_{t} -p_{r}\right)\label{eq24}
\end{eqnarray}
We plot (Fig.~3) the behavior of pressure anisotropy,
gravitational and hydrostatic forces in the stellar interior,
which clearly shows that the static equilibrium configurations do
exist due to the combined effect of pressure anisotropy,
gravitational and hydrostatic forces.

\begin{figure}
\centering
\resizebox{0.50\textwidth}{!}{
  \includegraphics{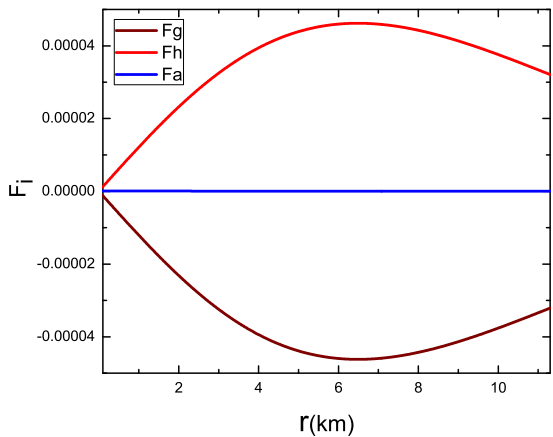}
} \caption{Behaviors of pressure anisotropy ($F_a$),
gravitational ($F_g$) and hydrostatic($F_h$) forces at the pulsar
 interior.} \label{fig:3}
\end{figure}

\subsection{Stability}
For a physically acceptable stellar model, one expects that the
speed of sound should be within the range  $0 \leq
v_s^2=(\frac{dp}{d\rho})\leq 1$ \cite{Herrera1992,Abreu2007}. In
our model, we plot the radial and transverse sound speeds in
Fig.~4 and observed that these parameters satisfies the
inequalities $0\leq v_{sr}^2 \leq1$ and $0\leq v_{st}^2 \leq 1$
everywhere within the stellar object. Since, $0\leq v_{sr}^2 \leq
1$ and $0\leq v_{st}^2 \leq 1$, therefore, $\mid v_{st}^2 -
v_{sr}^2 \mid \leq 1 $. In Fig.~4, we have plotted  $\mid
v_{st}^2 - v_{sr}^2 \mid$ for such verification. These results
shows that our dark matter admixed pulsars model is stable.

In our stellar model, the adiabatic index($\gamma$) satisfies the
inequality $\gamma = \frac{\rho+p_{r}}{p_{r}}
\frac{dp_{r}}{d\rho} > \frac{4}{3}$ everywhere within the stellar
interior (Fig.~4) which also verifies the dynamical stability of
our stellar model in presence of thermal radiation. This type of
stability executed by several author like Chandrasekhar
\cite{Chandrasekhar1964}, Bardeen et al. \cite{Bardeen1966},
Knutsen \cite{Knutsen1988}, Mak and Harko \cite{Harko2013}
gradually in their work.

\begin{figure*}
  \includegraphics[width=0.50\textwidth]{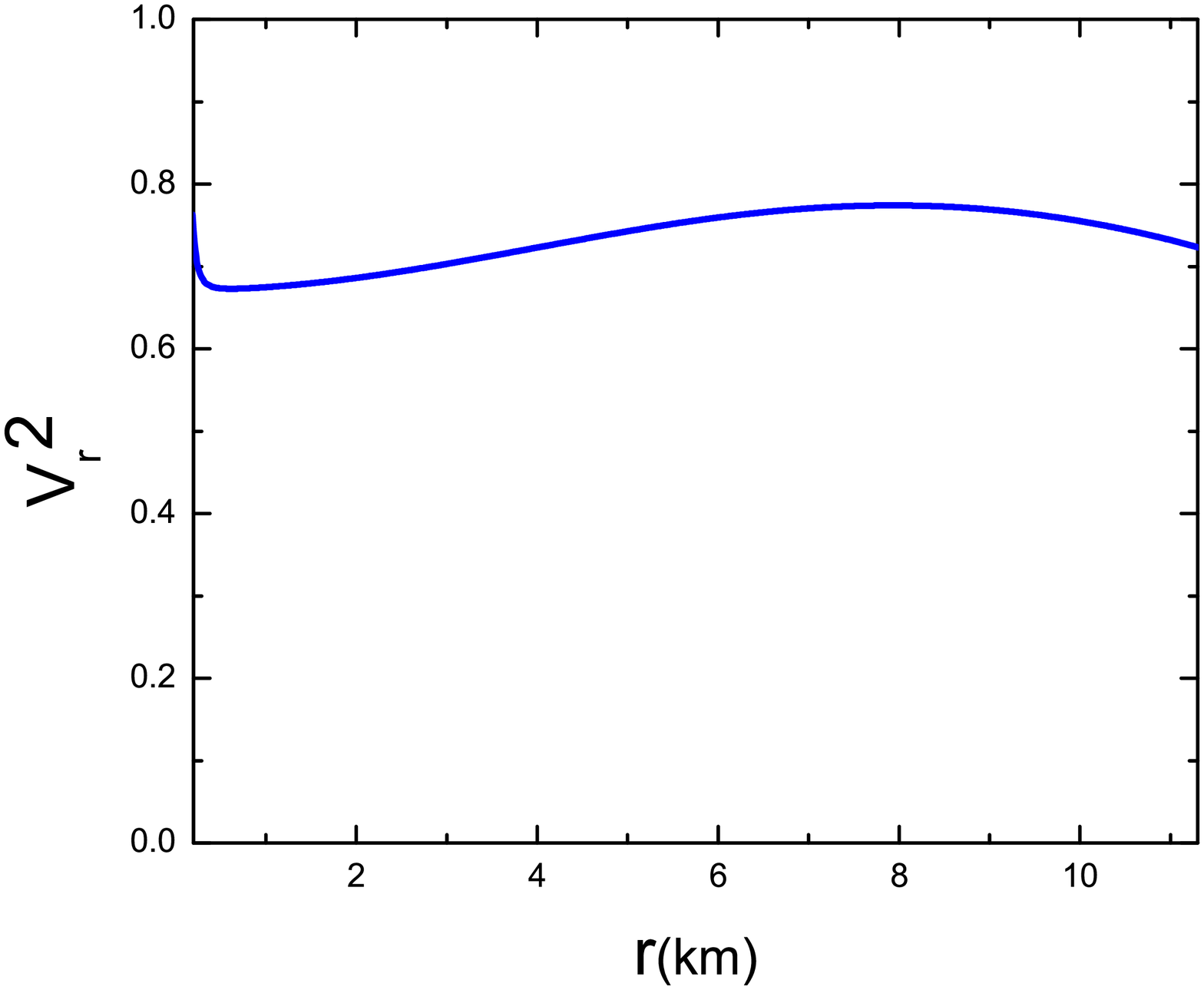}
  \includegraphics[width=0.50\textwidth]{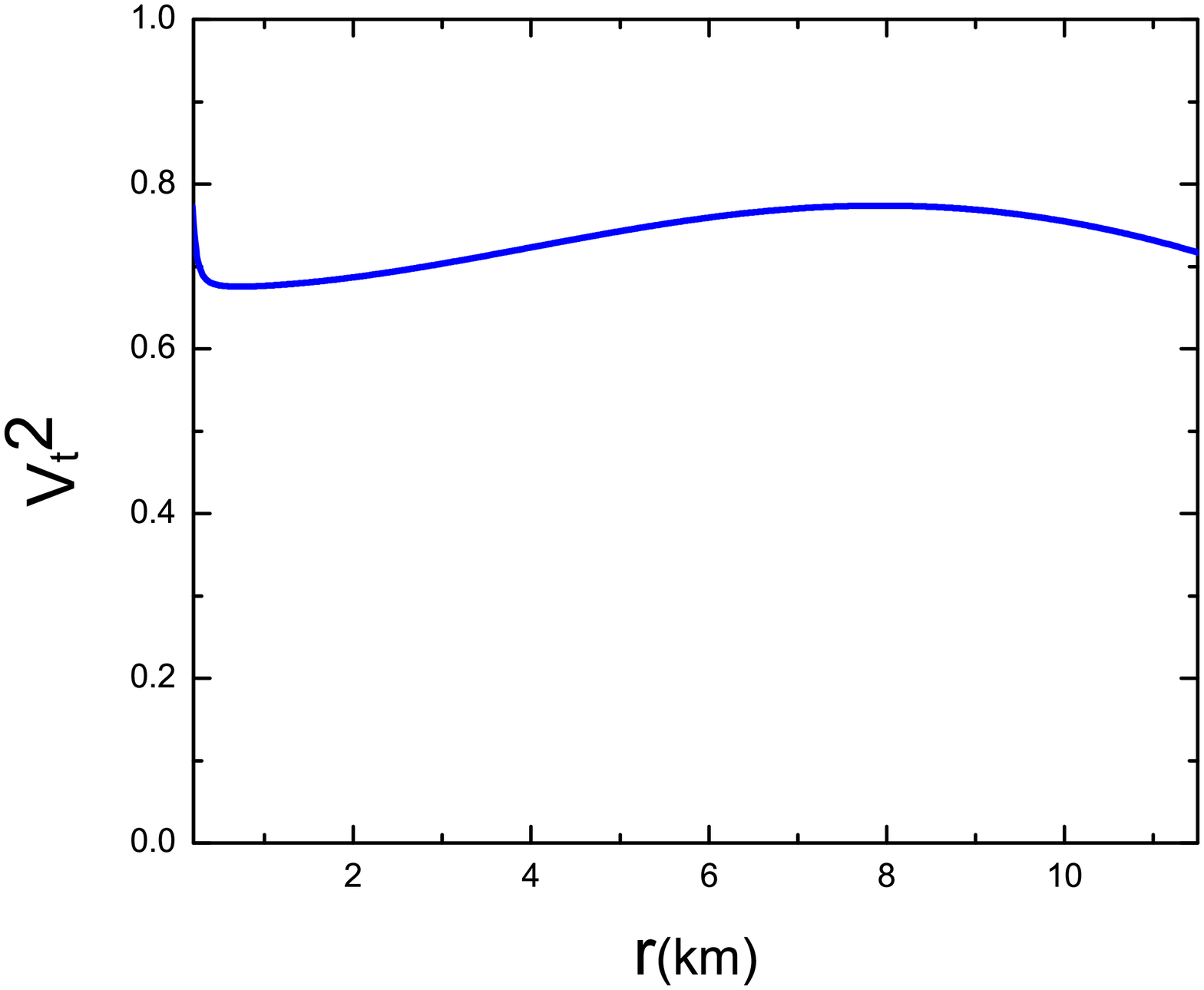}\\
   \includegraphics[width=0.50\textwidth]{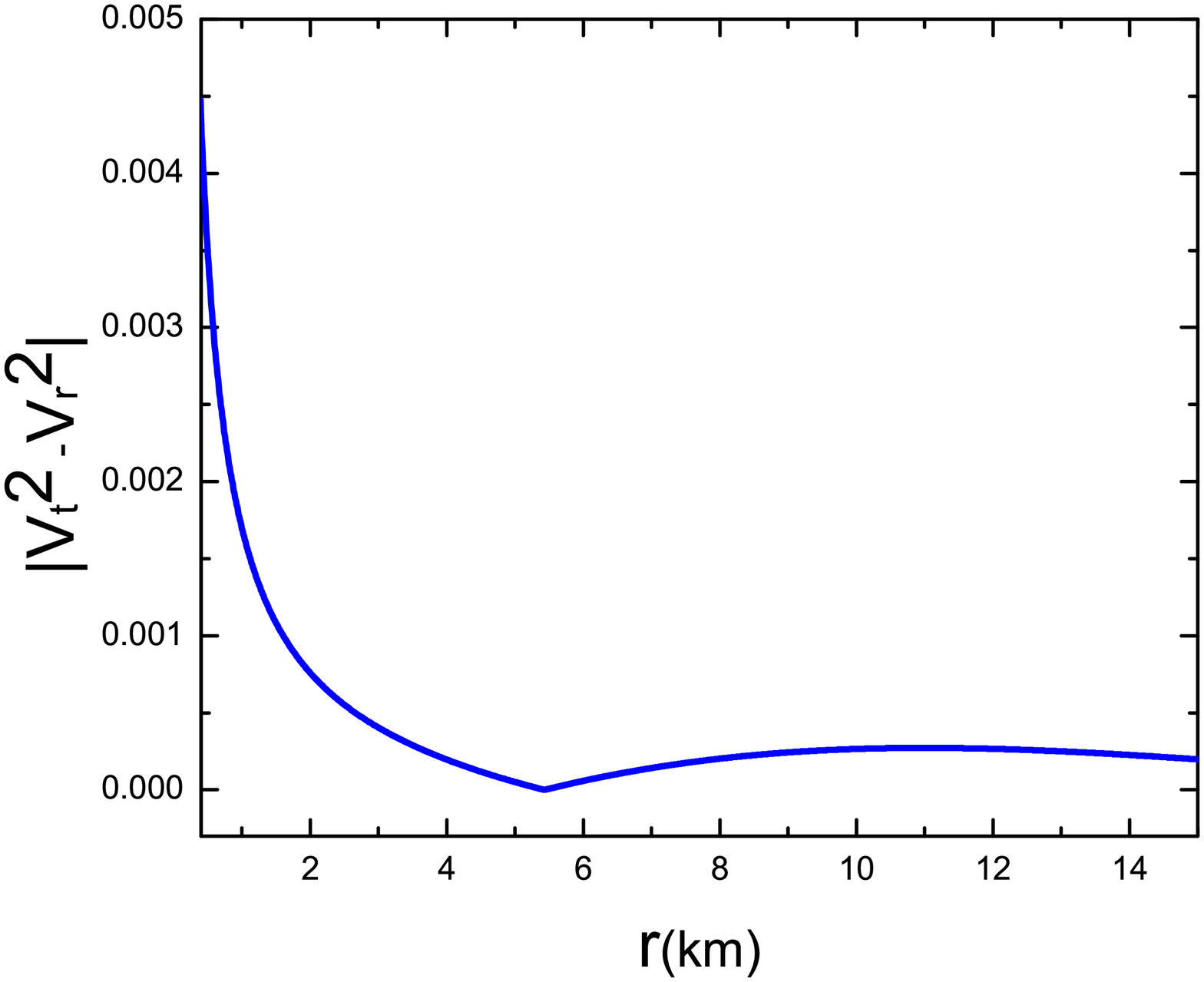}
  \includegraphics[width=0.50\textwidth]{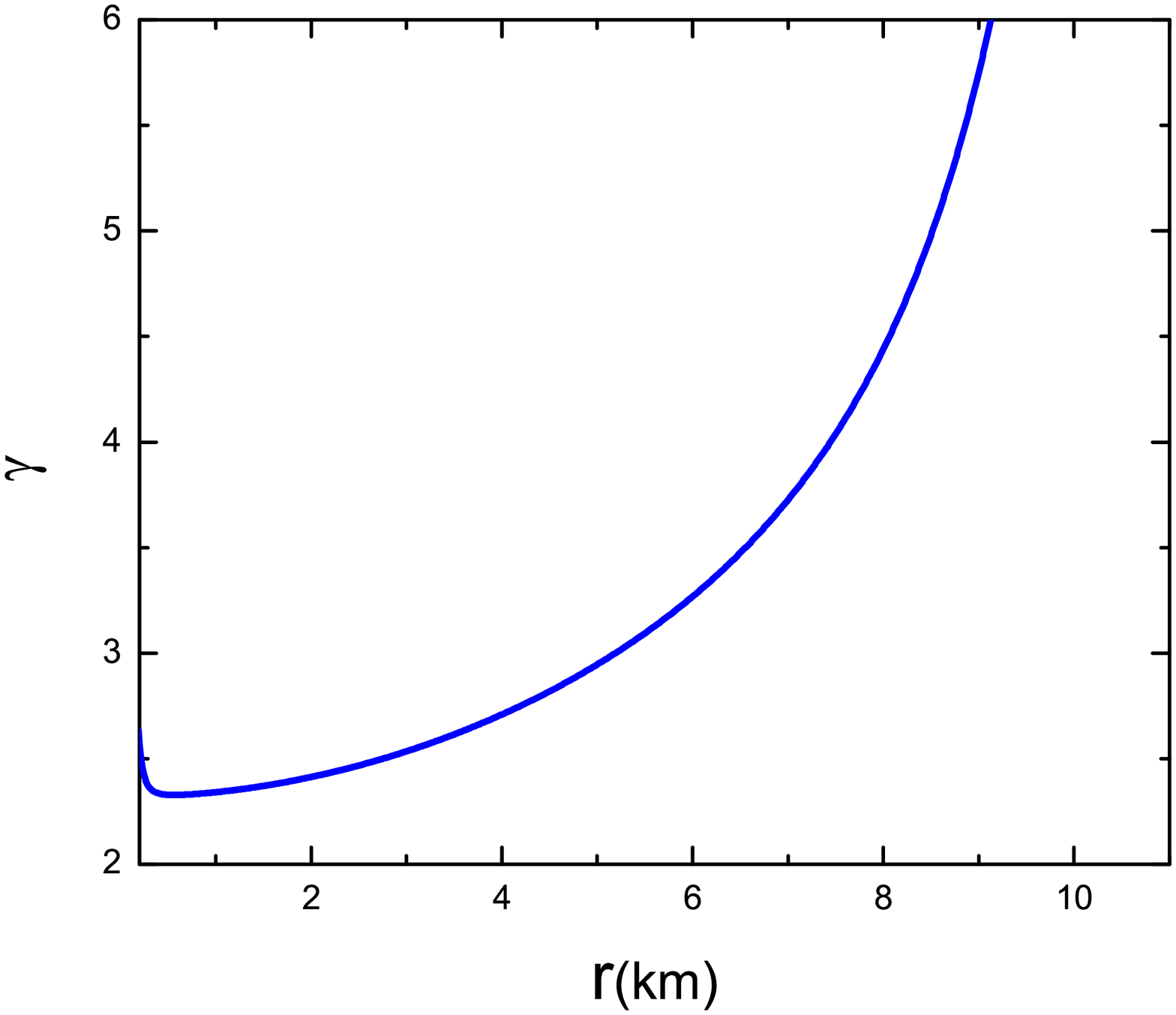}
\caption{Sound speed ($v^2$) - radius ($r$) and adiabatic index
($\gamma$) - radius ($r$) variation at the pulsar  interior.}
\label{fig:4}
\end{figure*}

\subsection{Matching Conditions}
The interior metric of the stellar body should match with the
Schwarzschild exterior metric at the boundary ($r=b$).
\begin{equation}
ds^2 = - \left(1-\frac{2M}{r}\right)dt^2 +  \left(1-\frac{2M}{r}\right)^{-1}dr^2 +r^2
d\Omega^2   \label{eq1}
\end{equation}
Assuming the continuity of the metric functions $g_{tt}, ~~g_{rr}$ and $\frac{\partial g_{tt}}{\partial r}$ at the boundary, we get
Continuity of the metric function across the boundary yields the compactification factor as
\begin{equation}
\frac{M}{b} =  \frac{1}{2}\left[\frac{3ab^2\left(1+C(1+4ab^2)^{-\frac{1}{2}}\right)}{2\left(1+ab^2\right)}\right]
\end{equation}

\subsection{Mass-Radius relation and Surface redshift}
According to Buchdahl \cite{Buchdahl1959}, maximum allowable mass-radius ratio for a static spherically symmetric perfect fluid sphere should be $\frac{ Mass}{Radius} < \frac{4}{9}$.
We have calculated the gravitational mass (M) as
\begin{equation}
\label{eq34}
 M=4\pi\int^{b}_{0} \rho_{eff}~~ r^2 dr = \frac{3ab^3\left(1+C(1+4ab^2)^{-\frac{1}{2}}\right)}{4\left(1+ab^2\right)}
\end{equation}
where $b$ is the radius of the pulsar.
\begin{figure}
\centering
\resizebox{0.50\textwidth}{!}{
  \includegraphics{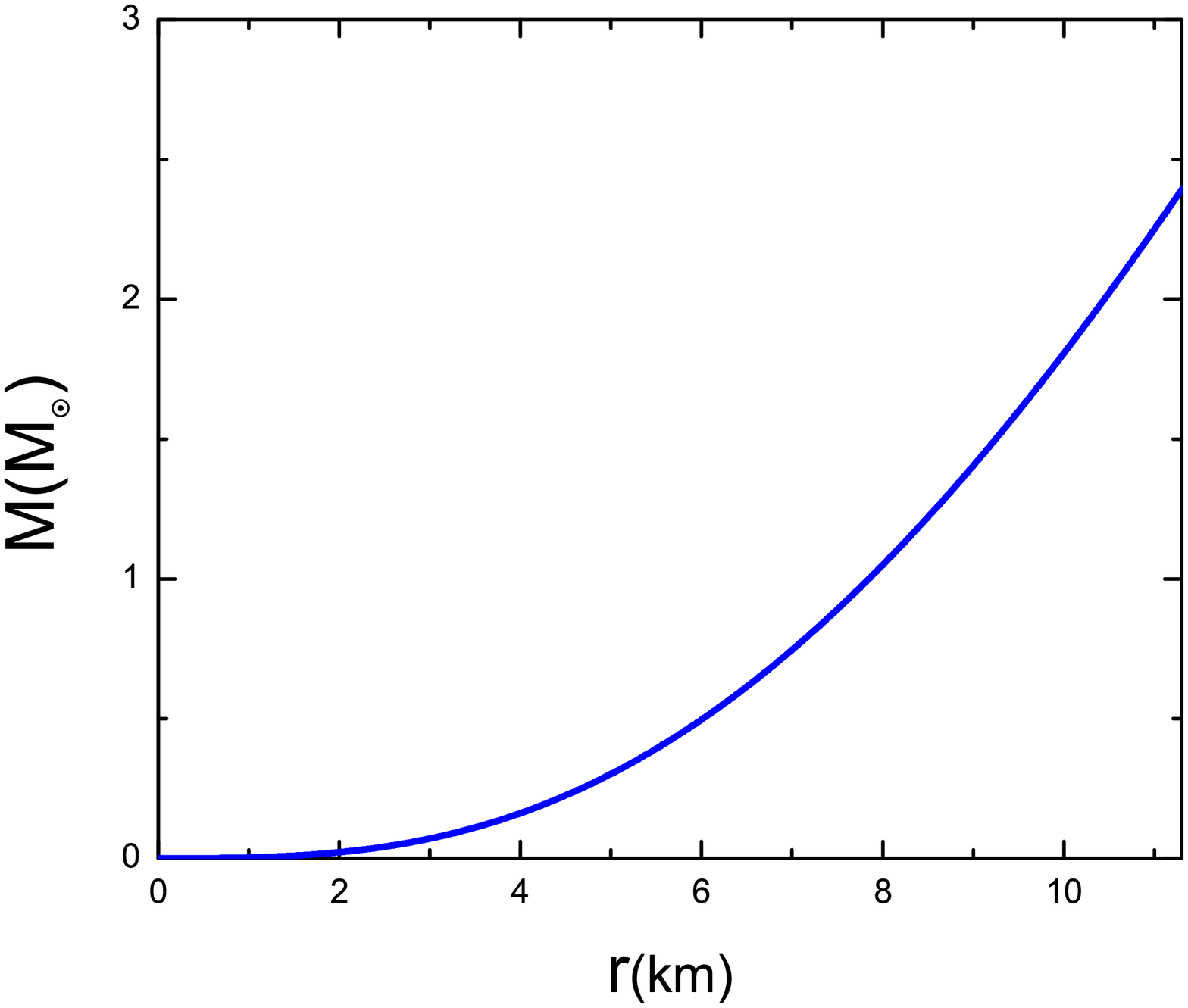}
}
\caption{Mass function M(r) variation at the pulsar interior.}
\label{fig:5}
\end{figure}
Therefore, the compactness (u) of the pulsar can be written as
\begin{equation}
u = \frac{M}{b} =  \frac{1}{2}\left[\frac{3ab^2\left(1+C(1+4ab^2)^{-\frac{1}{2}}\right)}{2\left(1+ab^2\right)}\right]
\end{equation}
The variation of mass function and compactness of the pulsar are
shown in Fig.~5 and Fig.~6 respectively. The surface redshift
($Z_s$) corresponding to the above compactness ($u$) can be
written as
\begin{eqnarray}
 Z_s= \left[ 1-2u \right]^{-\frac{1}{2}}-1 \\
 Z_s= -1+\frac{1}{\sqrt{1-\frac{3ab^2\left(1+\frac{C}{\sqrt{1+4ab^2}}\right)}{2+2ab^2}}}
\end{eqnarray}
\begin{figure}
\centering
\resizebox{0.50\textwidth}{!}{
  \includegraphics{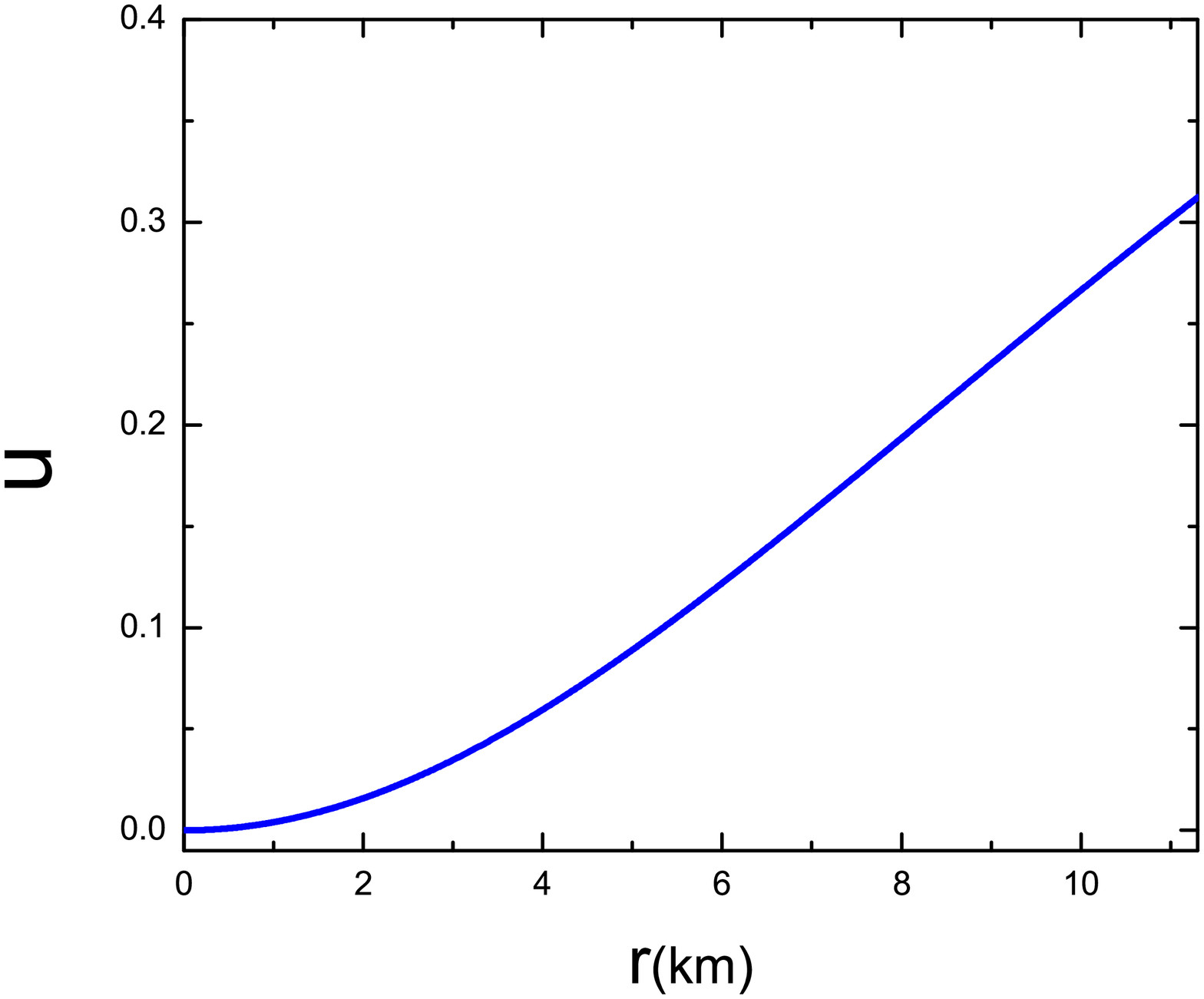}
} \caption{Variation of the compactness (u) at the  pulsar
interior.} \label{fig:6}
\end{figure}

\begin{figure}
\centering
\resizebox{0.50\textwidth}{!}{
  \includegraphics{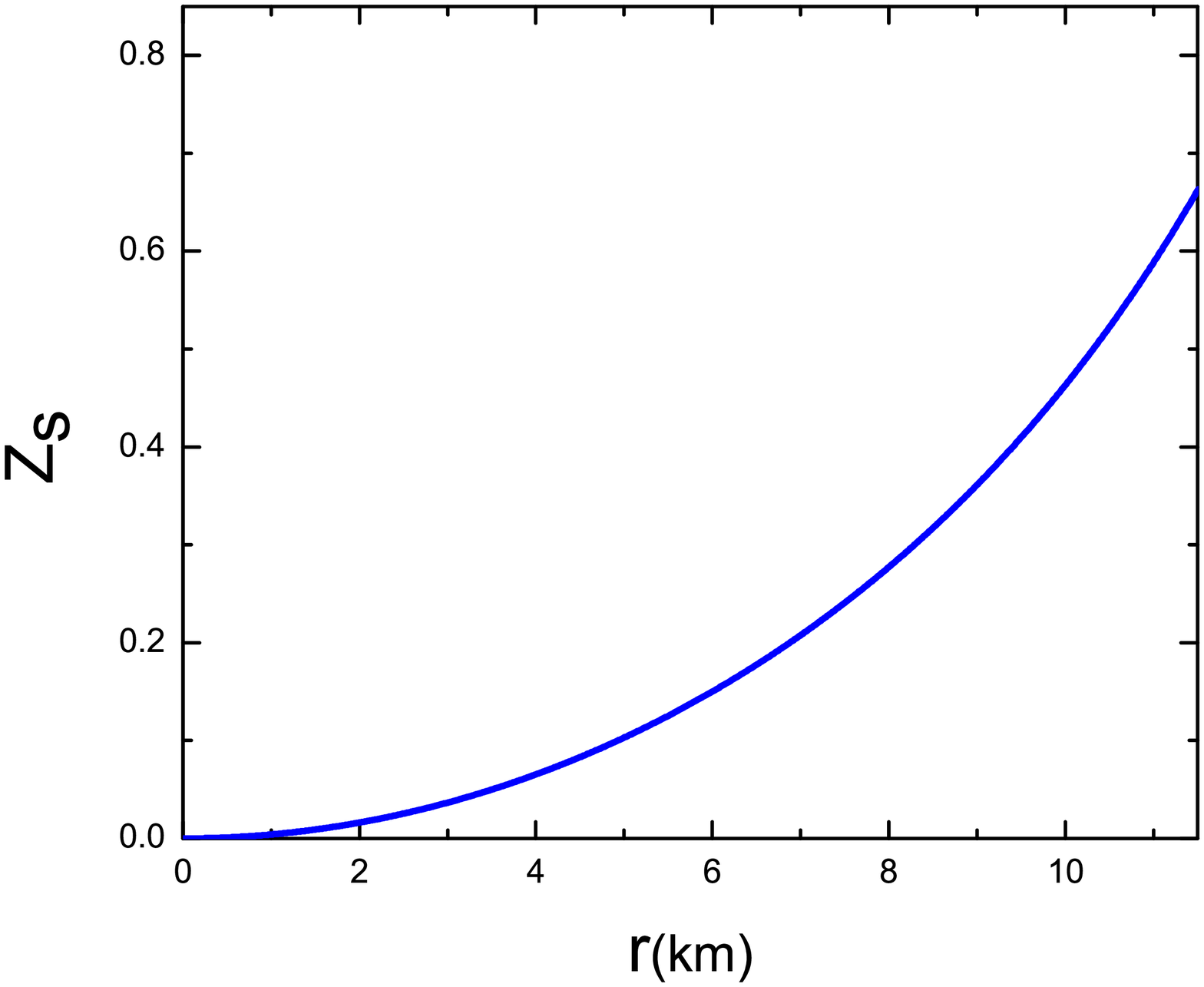}
} \caption{Variation of the surface redshift ($Z_s$)  at the
pulsar interior.} \label{fig:7}
\end{figure}

Therefore from Fig.~7, the maximum surface redshift for the different pulsars can be obtained easily. The radii, compactness and surface redshift of the different pulsars are evaluated from Fig.~8, Eq.~(18) and Eq.~(20) and a comparative analysis has been done in Table~1.

\begin{table*}
\caption{Evaluated parameters for pulsars}
\label{tab:1}       
\begin{tabular}{lllll}
\hline\noalign{\smallskip}
Star & Observed Mass($M_{\odot}$) & Radius from Model(in km) & Compactness from Model & Redshift from Model \\
\noalign{\smallskip}\hline\noalign{\smallskip}
PSR J1748-2021B & 2.74 $\pm$ 0.21 & 12.005 $\pm$ 0.425 & 0.336 $\pm$ 0.014 & 0.750 $\pm$ 0.075 \\
PSR J1911-5958A & $1.40^{+0.16}_{-0.10}$ & 8.72$\leq$ R $\leq$ 9.4 & 0.220$\leq$ u $\leq$ 0.245 & 0.336$\leq$ $Z_s$ $\leq$ 0.340 \\
PSR J1750-37A & $1.26^{+0.39}_{-0.36}$ & 7.53$\leq$ R $\leq$ 9.62  & 0.176$\leq$ u $\leq$ 0.253 & 0.243$\leq$ $Z_s$ $\leq$ 0.422 \\
PSR B1802-07 & $1.26^{+0.08}_{-0.17}$ & 8.13$\leq$ R $\leq$ 8.83 & 0.198$\leq$ u $\leq$ 0.224 & 0.287$\leq$ $Z_s$ $\leq$ 0.346 \\
\noalign{\smallskip}\hline
\end{tabular}
\end{table*}

\begin{figure*}
  \includegraphics[width=0.50\textwidth]{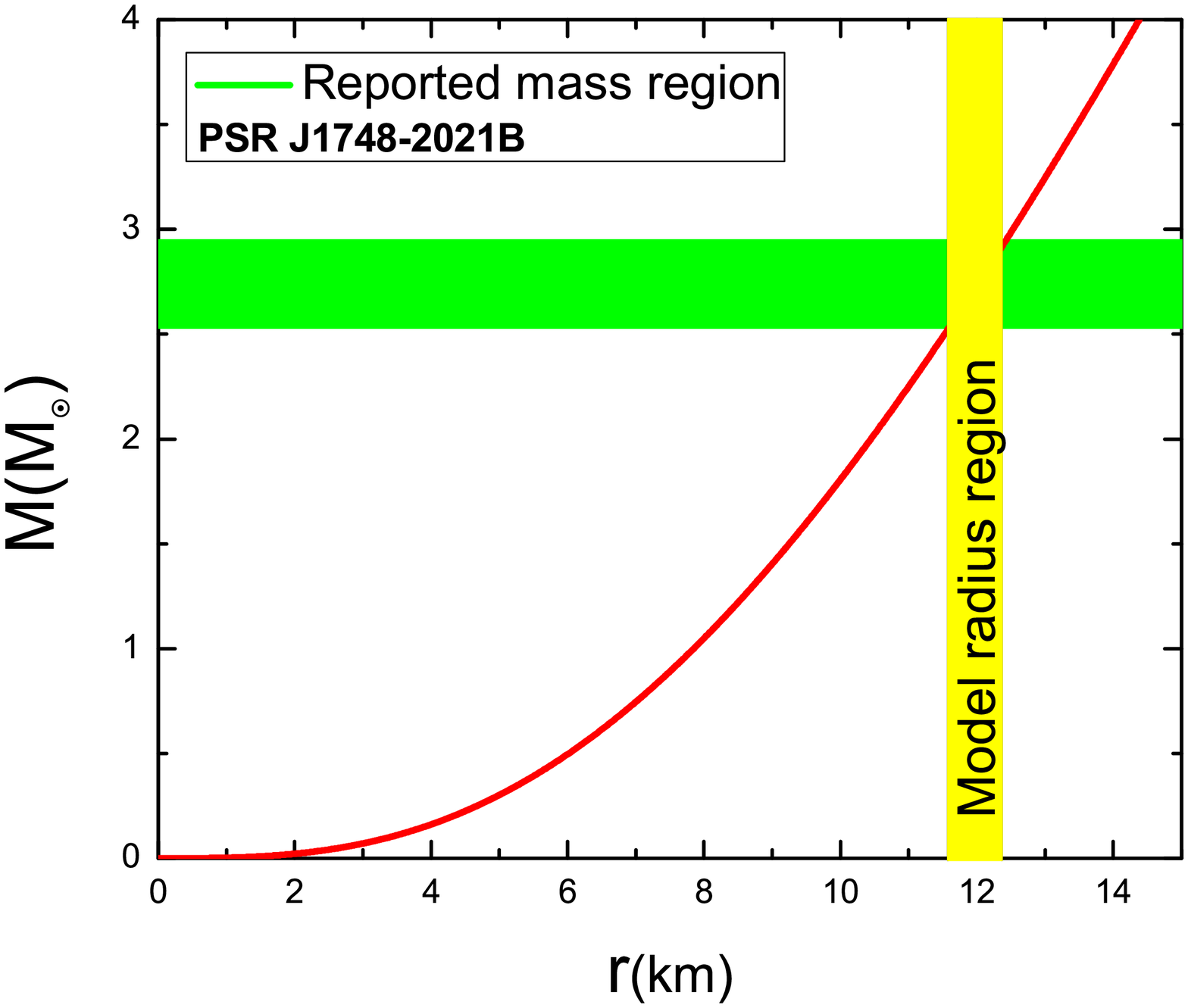}
  \includegraphics[width=0.50\textwidth]{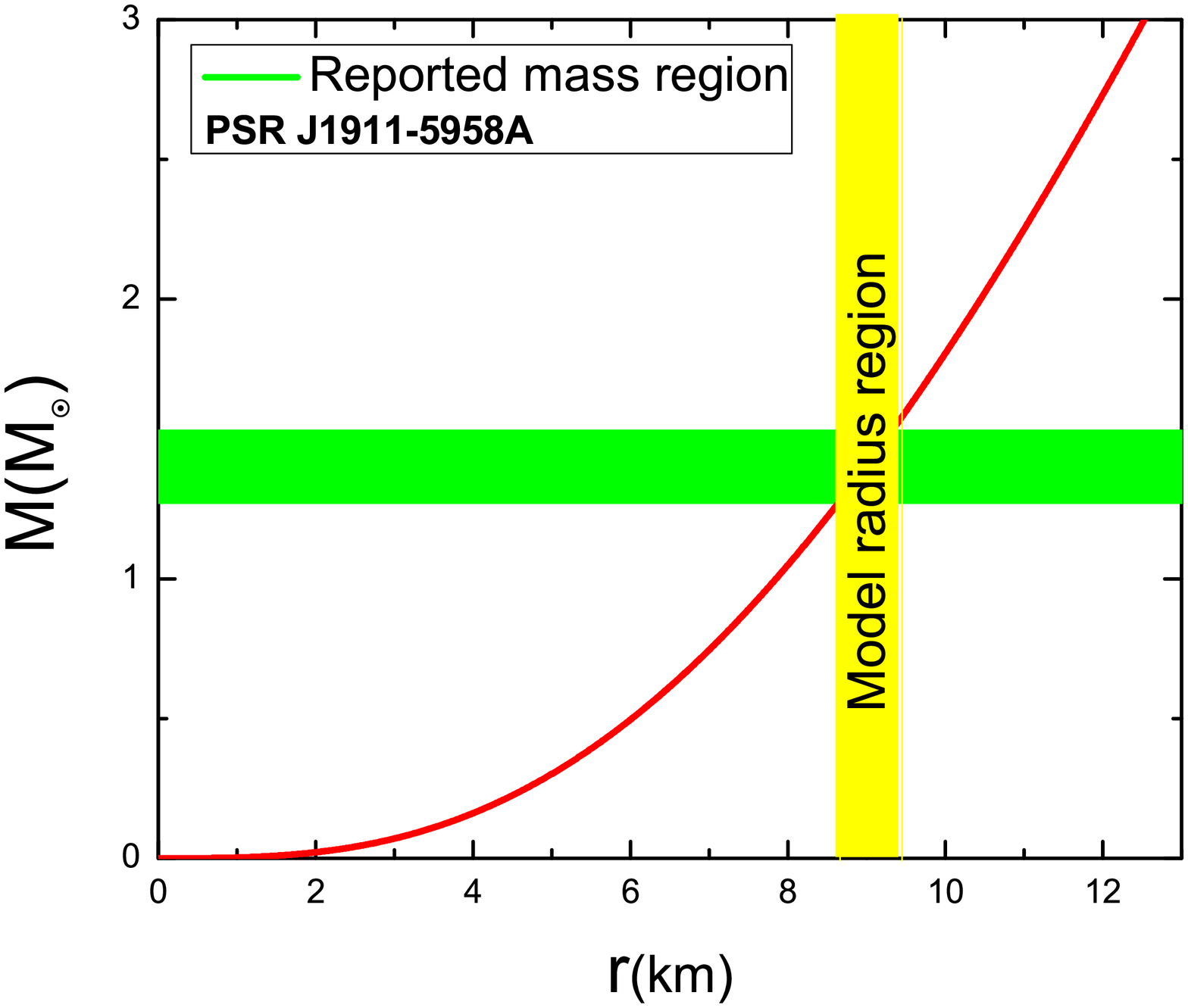}\\
   \includegraphics[width=0.50\textwidth]{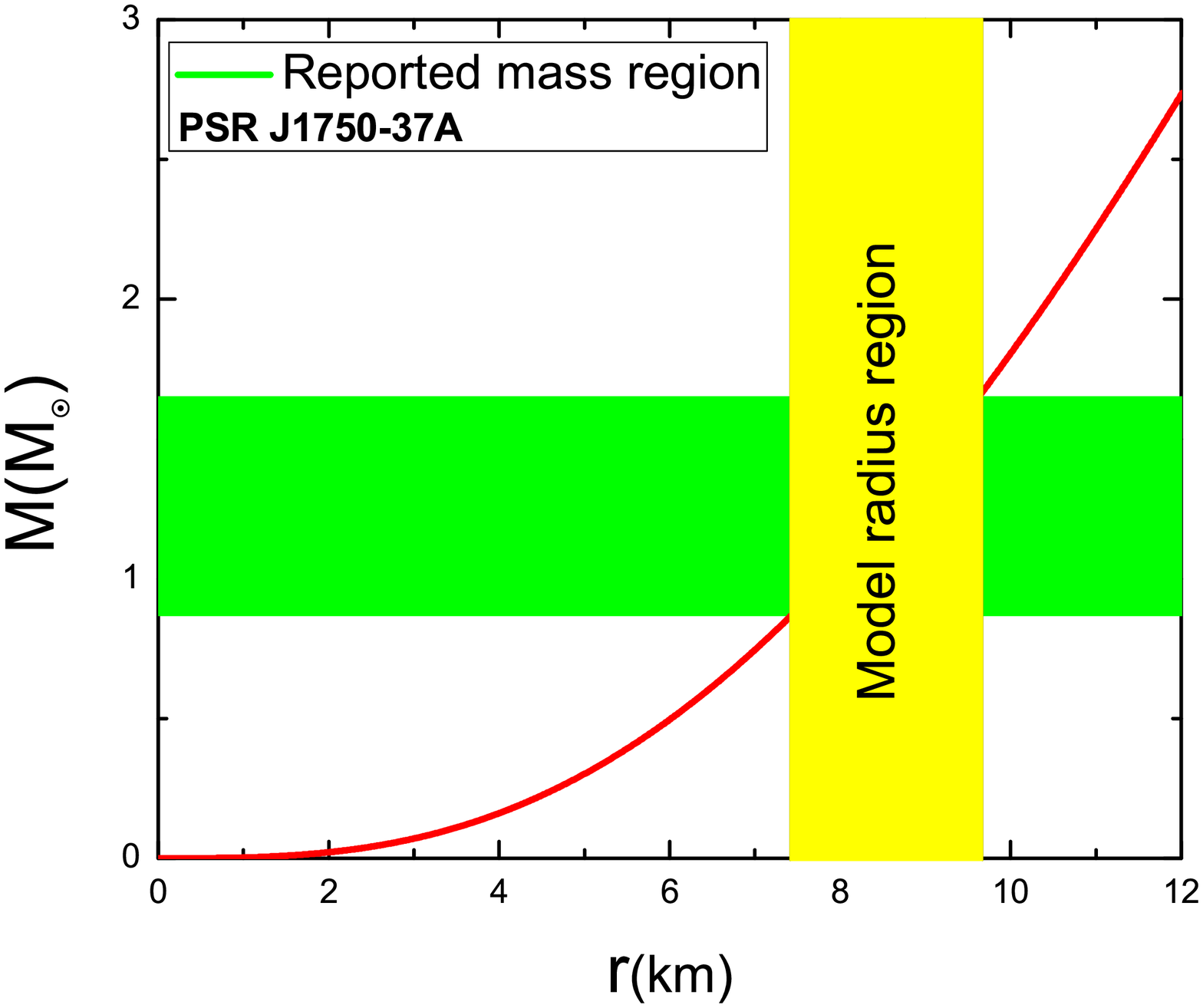}
  \includegraphics[width=0.50\textwidth]{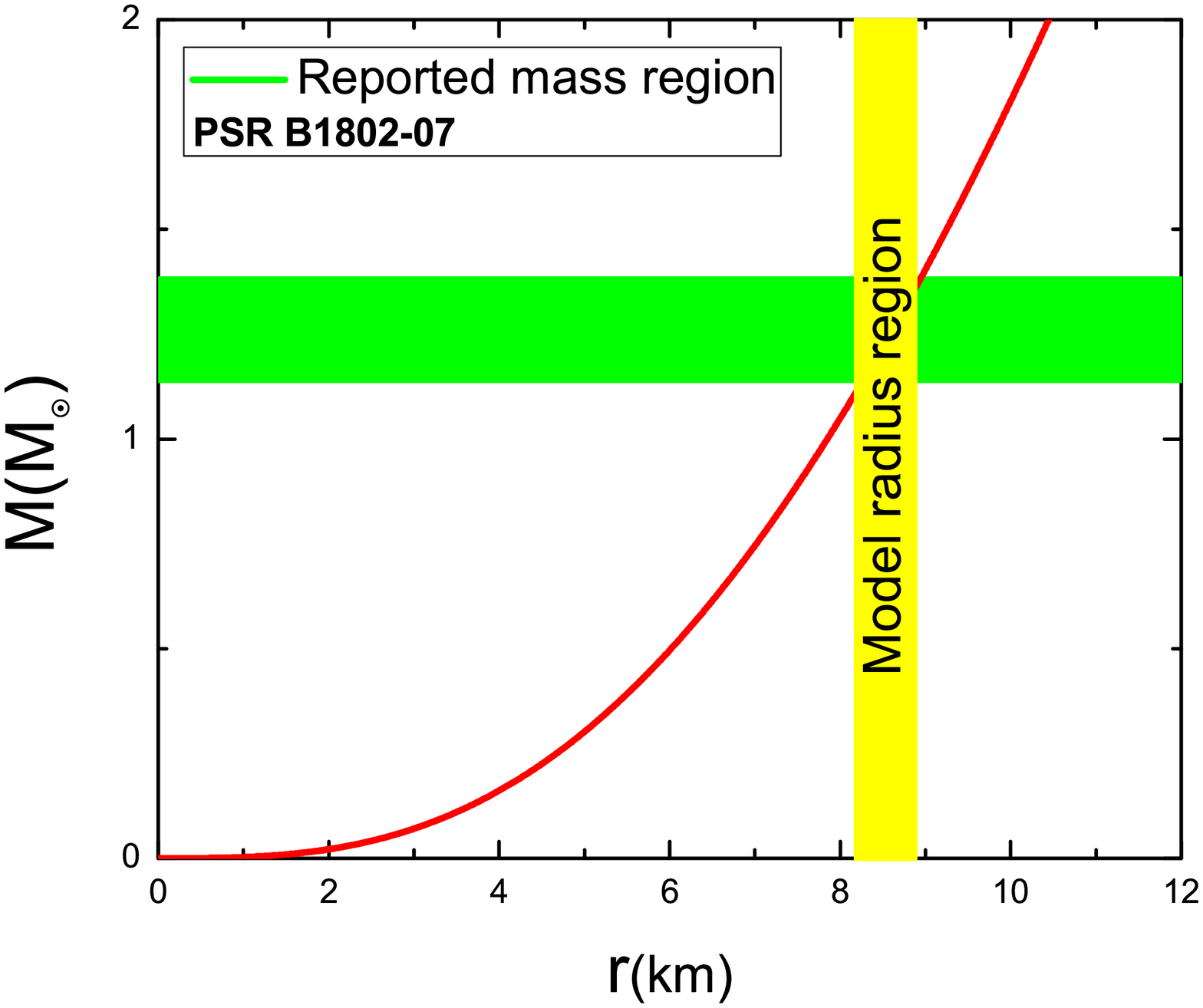}
\caption{Probable radii of Pulsars.} \label{fig:8}
\end{figure*}

\section{Discussion and Concluding Remarks}
\label{sec:3} In the present work, we investigate the nature of
the pulsars present in different galaxy by taking the Heintzmann
IIa metric. In general pulsars, due to their high density,
becomes anisotropic in nature. That's why we consider the
anisotropic behaviour of the pulsar to make more generalized
model.Also, we have been motivated by the previous articles on
dark matter neutron star reported in well known journals
\cite{Li2012b,Leung2011,Panotopoulos2017b,Sandin2009,Goldman2013,Leung2013,Mukhopadhyay2017,Rezaei2017,Rezaei2018}.
But our approach is quite different from these articles.
Conventionally, the mass-radius curve of compact stars are
calculated under a given equation of state for various values of
central density; by a given value of the central density, the
mass and radius of a compact star are fixed.  According to our
model, different pulsars are depends on the same
 parameter values $K=10^{-8}$ ( as observed  $K$ varies from $10^{-7}$
to  $10^{-8}$), $a=0.003 km^{-2}$, $C=0.8$ and $m=10^{-6}$,
 such that it obeys the standard star model condition ( i.e. pressure drops to zero at the boundary etc.).
Consequently, they have the same central density and the same
equation of state. Therefore, interestingly, if we starts from
the centre with a certain central density, the model of a compact
star can be determined by stopping at any radius where pressure
becomes zero. We think this model will give new dimension to
study of compact stars.

As the structure of the pulsar are still not known,we have
considered a two-fluid model assuming that the pulsars are made
of ordinary matter admixed with dark matter having a
characteristic parameter $\rho_{d}$.Contribution of dark matter
comes from the fitting of the rotation curves of the SPARC sample
of galaxies\cite{Lelli2016}. For this we have investigated the
dark matter based on the Singular Isothermal Sphere (SIS) dark
matter density profile in the galactic halo region.\\

Main motivation of the present article is to study the pulsars(
consists of dark matter with ordinary matter) presents in
different galaxies, namely, PSR J1748-2021B in NGC 6440B, PSR
J1911-5958A in NGC 6752, PSR B1802-07 in NGC 6539, PSR J1750-37A in NGC 6441.\\

Successfully, we find an analytical solution to the fluid sphere
which are quite interesting in connection to several physical
features, which are as follows:
\begin{enumerate}
\item In our model, density and pressure at the interior of the pulsar are well behaved (Fig.~1). Pressure and density are
both maximum at the centre and monotonically decreasing towards
the boundary. Here, we assume the values of constants ($a$, $C$)
in the metric and m, $K=10^{-8}$ ( as observed  $K$ varies from
$10^{-7}$ to  $10^{-8}$) in such a way that pressure must vanish
at the boundary. Our pulsar model satisfies all the energy
conditions, TOV equation and Herrera's stability condition
\cite{Herrera1992}. It is also stable with respect to
infinitesimal radial thermal perturbations. From the mass
function (Eq.~17), all desired interior features of a pulsar can
be evaluated which satisfies Buchdahl mass-radius relation
($\frac{ 2M}{R} < \frac{8}{9}$) (Figs.~5 and 6). The surface
redshift of the pulsar are found within the standard value
($Z_{s}\leq 0.85$) which is satisfactory (Fig.~7)
\cite{Haensel2000}.

\item From our mass function graphs fig.~8, Eq.~(18) and Eq.~(20), we obtain the radii, compactness and surface red-shift of four pulsars in the various
galaxy namely: PSR J1748-2021B in NGC 6440B, PSR J1911-5958A in
NGC 6752, PSR B1802-07 in NGC 6539 and PSR J1750-37A in NGC 6441.
The detail comparison chart are shown in Table~3.
\end{enumerate}

It is to be mentioned here that the K values for both dwarf and
spiral galaxies are of the order of $10^{-7}$ to $10^{-8}$ and
that value of K comes from $v_{halo}^2$,the observational data of
galactic rotational curve. Moreover, this K has an important role
to the density distribution of the dark matter haloes. In our
calculation, we consider the value of K as $10^{-8}$ (i.e.
contribution comes from Galactic Rotation Curve data).




In conclusion, we can say that incorporation of dark matter with
the real matter one can describe the well-known pulsars (e.g. PSR
J1748-2021B in NGC 6440B, PSR J1911-5958A in NGC 6752, PSR
B1802-07 in NGC 6539 and PSR J1750-37A in NGC 6441 etc.) in a
good manner in all respects. Therefore, we conclude that there is
every possibility of existence of dark matter admixed with
ordinary matter in the above mentioned pulsars.

\begin{acknowledgements}
I acknowledge H. Haghi and A. Ghari of IASBS, Iran for providing
the data of $v_{halo}^2$ which was calculated from the fitting of
the rotation curves of the SPARC sample of galaxies. MK would
like to thank IUCAA, Pune, India for providing research
facilities and warm hospitality under Visiting Associateship
where a part of this work was carried out.
\end{acknowledgements}

\end{document}